\documentclass[conference]{IEEEtran}
\IEEEoverridecommandlockouts
\usepackage{bm}
\usepackage{cite}
\usepackage{amsmath,amssymb,amsfonts}
\usepackage{algorithmic}
\usepackage{graphicx}
\usepackage{textcomp}
\usepackage{xcolor}
\usepackage{subfigure}

\def\BibTeX{{\rm B\kern-.05em{\sc i\kern-.025em b}\kern-.08emT\kern-.1667em\lower.7ex\hbox{E}\kern-.125emX}}
\newtheorem{assumption}{Assumption}

\newtheorem{corollary}{Corollary}

\newtheorem{theorem}{Theorem}

\begin{document}
\title{Over-the-Air Computation Empowered Federated Learning: A Joint Uplink-Downlink Design}
\author{\IEEEauthorblockN{Deyou Zhang, Ming Xiao, and Mikael Skoglund}
\IEEEauthorblockA{\textit{Division of Information Science and Engineering, KTH Royal Institute of Technology, Stockholm, Sweden} \\
email: \{deyou, mingx, skoglund\}@kth.se}
}

\maketitle

\begin{abstract}
In this paper, we investigate the communication designs of over-the-air computation (AirComp) empowered federated learning (FL) systems considering uplink model aggregation and downlink model dissemination jointly. We first derive an upper bound on the expected difference between the training loss and the optimal loss, which reveals that optimizing the FL performance is equivalent to minimizing the distortion in the received global gradient vector at each edge node. As such, we jointly optimize each edge node transmit and receive equalization coefficients along with the edge server forwarding matrix to minimize the maximum gradient distortion across all edge nodes. We further utilize the MNIST dataset to evaluate the performance of the considered FL system in the context of the handwritten digit recognition task. Experiment results show that deploying multiple antennas at the edge server significantly reduces the distortion in the received global gradient vector, leading to a notable improvement in recognition accuracy compared to the single antenna case.
\end{abstract}


\IEEEpeerreviewmaketitle

\section{Introduction}
With the widespread deployment of 5G communication networks, there has been growing interest in exploring 6G communications in both academia and industry \cite{Ming-6G, Akyildiz-6G, Letaief-CM}. It is widely anticipated that 6G communications will rely on ubiquitous artificial intelligence to achieve data-driven machine learning (ML) solutions in large-dimensional and heterogeneous networks \cite{Letaief-CM}. However, traditional ML techniques typically require a centralized data collection process, which consumes substantial communication and computation resources and often leads to severe latency. Additionally, such a centralized data collection process can also raise privacy and security concerns, particularly when dealing with sensitive data.

As an emerging distributed ML approach, federated learning (FL) provides a new paradigm to cope with these concerns \cite{FL-Google}. In typical FL frameworks, it is unnecessary for smart edge nodes to reveal their local data to the edge server. Instead, FL repeatedly executes the following two processes. 1) Model aggregation: edge nodes upload their respective local model parameters\footnote{Local model parameters of each edge node are computed based on its received global model parameters and its own dataset, as detailed in Section \ref{Section-FLPreliminary}. Moreover, the computation of local model parameters is often referred to as local training.} to the edge server, which averages over those parameters to obtain global model parameters. 2) Model dissemination: the edge server broadcasts global model parameters to edge nodes for the next local training. Since only model parameters instead of raw data are sent to the edge server, FL is capable of achieving privacy protection and relieving communication burdens.

Despite the advantages of FL, uploading local model parameters via traditional orthogonal multiple access (OMA) protocols is resource-demanding, and it has become a bottleneck for implementing FL in practice. In light of this, several recent works proposed to optimize resource allocation among edge nodes to enhance the communication efficiency in model uploading \cite{MingzheChen-TWC, HaoChen-IoTJ}. Though with merits, those works did not exploit the waveform-superposition property of multiple-access channels and thus did not fully unleash the benefits of wireless communications. As an alternative, over-the-air computation (AirComp) empowered model aggregation has recently emerged \cite{GuangxuZhu-TWC, Deniz-TSP, YuanmingShi-TWC}.

The first AirComp-empowered model aggregation research appeared in \cite{GuangxuZhu-TWC}, demonstrating that AirComp could significantly reduce the model uploading latency compared to its OMA counterpart. Meanwhile, gradient sparsification and compression methods were investigated in \cite{Deniz-TSP} to further alleviate the uplink communication burden. As preliminary works, only single-input single-output (SISO) configuration was considered in those literatures \cite{GuangxuZhu-TWC, Deniz-TSP}. To leverage the benefit of multi-antenna technology, authors in \cite{YuanmingShi-TWC} focused on the single-input multiple-output (SIMO) configuration and proposed to jointly optimize edge node selection and the edge server receive beamforming vector to control the communication errors in model aggregation. While in \cite{YongmingHuang-JSAC}, both multiple-input single-output (MISO) and multiple-input multiple-output (MIMO) configurations were considered. Moreover, to better adapt to the wireless fading channel, authors in \cite{YongmingHuang-JSAC} introduced local learning rates and, based on which, proposed a modified federated averaging algorithm. Nonetheless, the beamforming designs in \cite{YuanmingShi-TWC, YongmingHuang-JSAC} only considered the model aggregation phase and ignored the model dissemination phase. In realistic FL systems, the two phases are intertwined, and the quality of model transmission in one phase affects the other one. Consequently, the two phases need to be considered jointly such that new beamforming designs remain to be explored \cite{ShuaiWang-TWC}.

In this paper, we focus on a typical wireless FL system consisting of one multi-antenna edge server and multiple single-antenna edge nodes. AirComp is employed for model aggregation, and the popular uniform-forcing design \cite{LiChen} is adopted to recover a noisy version of the global gradient vector at each edge node after model dissemination. Different from existing works, we investigate the communication designs of AirComp-empowered FL considering uplink model aggregation and downlink model dissemination jointly.

Specifically, we first derive an upper bound on the expected difference between the training loss and the optimal loss, which reveals that optimizing the FL performance is equivalent to minimizing the mean squared error (MSE) between the target global gradient vector and the received one at each edge node. Then, we seek to jointly optimize each edge node transmit and receive equalization coefficients along with the edge server forwarding matrix to minimize the maximum MSE across all edge nodes, which is a highly non-convex optimization problem. Considering the uniform-forcing design criterion, we prove that the edge server forwarding matrix must be a rank-one matrix. Leveraging this property, we decompose the original problem into two subproblems and optimize the uplink and downlink transceiver designs separately. While each subproblem remains non-convex, we employ the matrix lifting technique to transform them into difference-of-convex (DC) programs, which can be efficiently solved using the successive convex approximation (SCA) technique. Furthermore, we utilize the MNIST dataset to evaluate the performance of the considered FL system in the context of the handwritten digit recognition task. Experiment results show that deploying multiple antennas at the edge server significantly reduces the MSE at each edge node and improves the recognition accuracy compared to the single antenna case.

Indeed, the literature on joint uplink-downlink communication design for FL was scarce \cite{CongShen-JSAC, ZhibinWang-JSAC, ChongZhang-JDU}. Specifically, authors in \cite{CongShen-JSAC} investigated the impact of different model quantization methods on FL learning performance considering both uplink and downlink transmission. In \cite{ZhibinWang-JSAC}, FL over a multi-cell network accounting for inter-cell interference in both uplink and downlink communication was explored. However, both of these works only considered the SISO configuration. The closest work to ours is \cite{ChongZhang-JDU}, where a multi-antenna edge server was also assumed. In this work, both uplink and downlink beamforming designs were considered but with different approaches.

Throughout this paper, we use regular, bold lowercase, and bold uppercase letters to denote scalars, vectors, and matrices, respectively; $\mathcal R$ and $\mathcal C$ to denote the real and complex number sets, respectively; $(\cdot)^T$ and $(\cdot)^H$ to denote the transpose and the conjugate transpose, respectively. We use $x_i$ to denote the $i$-th entry in $\mathbf x$; $\|\mathbf x\|$ to denote the $\ell_2$-norm of $\mathbf x$; $\rm{diag}(\mathbf x)$ to denote a diagonal matrix with its diagonal entries specified by $\mathbf x$. We use $|\mathcal D|$ to denote the cardinality of set $\mathcal D$; $\langle \mathbf A, \mathbf B \rangle$ to denote the inner product of $\mathbf A$ and $\mathbf B$. We use $\mathbf I$ to denote the identity matrix; $\mathcal {CN}(\bm \mu, \bm \Sigma)$ to denote the complex Gaussian distribution with mean $\bm \mu$ and covariance matrix $\bm \Sigma$; $\nabla$ to denote the gradient operator, and ${\mathbb E}$ to denote the expectation operator.

\section{System Model} \label{Section-SystemModel}
In this section, we first provide some preliminary knowledge about FL and then introduce the AirComp-empowered FL framework. Note that both uplink model aggregation and downlink model dissemination are considered for communication designs.

\subsection{FL Systems} \label{Section-FLPreliminary}
\begin{figure}
\vskip 2pt
\centering
\subfigure[Model aggregation]
{\includegraphics[width = 4.2cm]{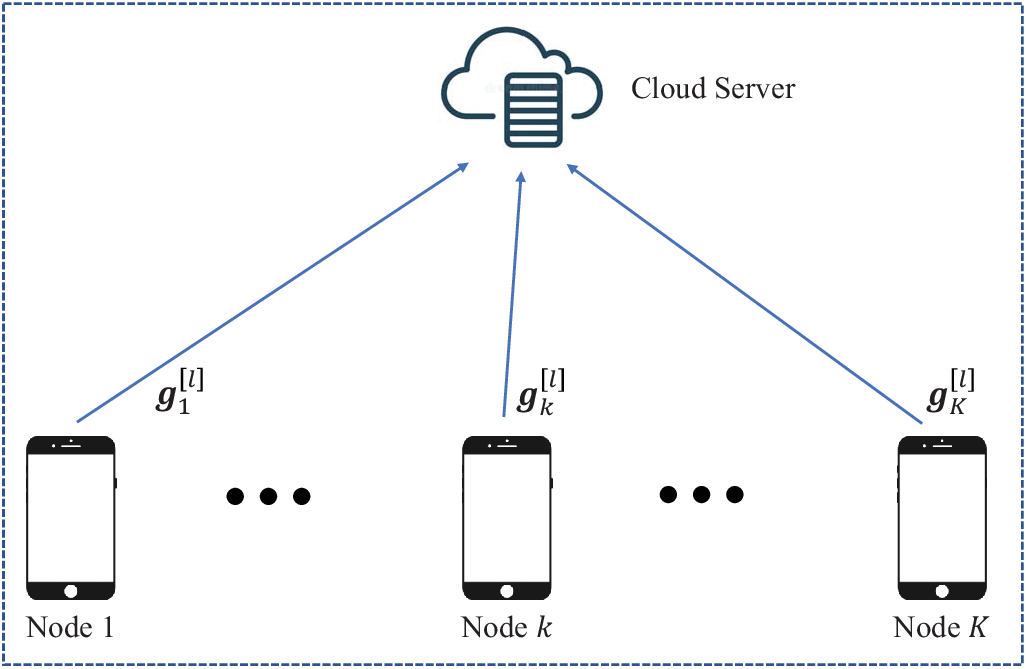}\label{MA}}
\subfigure[Model dissemination]
{\includegraphics[width = 4.2cm]{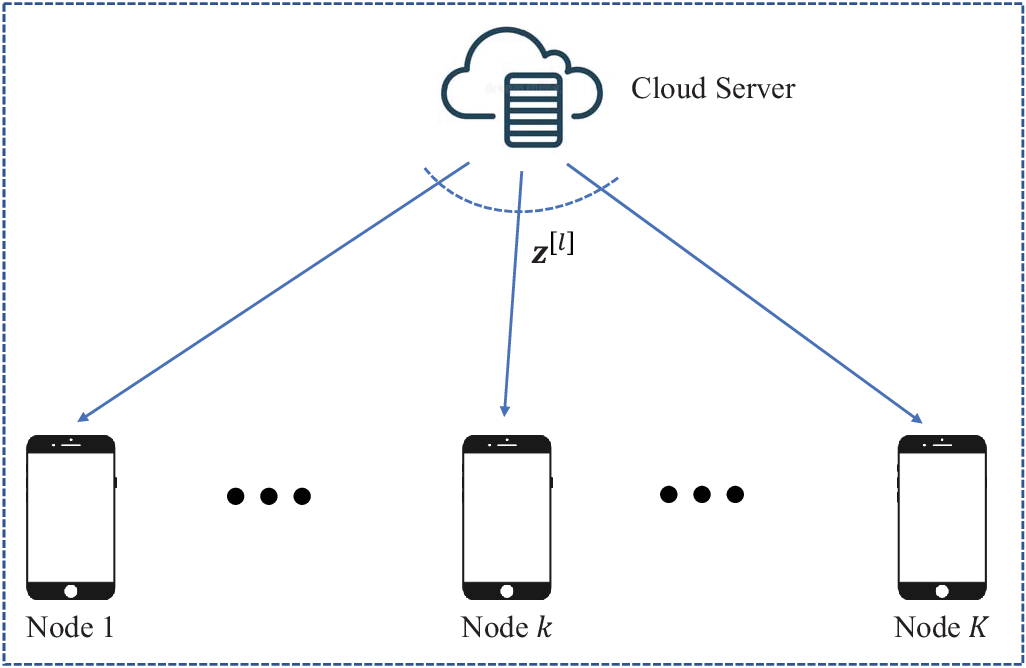}\label{MB}}
\caption{Illustration of the $l$-th FL training round.}\label{FL}
\vspace{-1em}
\end{figure}

As depicted in Fig. \ref{FL}, a typical FL system consists of an edge server and $K$ edge nodes. Edge node $k$, $\forall k \in {\cal K} \triangleq \{1, \cdots, K\}$, has a local dataset ${\cal D}_k$ that contains $D_k \triangleq |{\cal D}_k|$ labeled data samples, denoted by $\{\bm \xi_{k,1}, \zeta_{k,1}\}$, $\cdots, \{\bm \xi_{k,D_k}, \zeta_{k,D_k}\}$. Here, tuple $\{\bm \xi_{k,i}, \zeta_{k,i}\}$ denotes the $i$-th data sample in ${\cal D}_k$, consisting of a feature vector $\bm \xi_{k,i}$ and its corresponding ground-truth label $\zeta_{k,i}$, $\forall i \in \{1, \cdots, D_k\}$. The objective of FL is to seek a model parameter vector $\bm \theta \in {\cal R}^d$ that minimizes the following global loss function
\begin{equation}\label{Eq-GLF}
    F(\bm \theta) = \frac{1}{\sum\nolimits_{j=1}^K D_j} \sum\limits_{k=1}^K \sum\limits_{i=1}^{D_k} f(\bm \theta; \bm \xi_{k, i}, \zeta_{k,i}),
\end{equation}
in a distributed manner, where $f(\bm \theta; \bm \xi_{k, i}, \zeta_{k,i})$ is termed sample-wise loss function quantifying the misfit of $\bm \theta$ on the data sample $\{\bm \xi_{k, i}, \zeta_{k,i}\}$.

To this end, we follow \cite{FL-Google, MingzheChen-TWC, HaoChen-IoTJ, GuangxuZhu-TWC} and define the local loss function of $\bm \theta$ on ${\cal D}_k$, $\forall k \in {\cal K}$, as
\begin{equation}\label{LLF}
    F_k(\bm \theta) = \frac{1}{D_k} \sum\limits_{i=1}^{D_k} f(\bm \theta; \bm \xi_{k, i}, \zeta_{k,i}).
\end{equation} 
Then, the global loss function in \eqref{Eq-GLF} can be rewritten as
\begin{equation}\label{Eq-GLF2}
    F(\bm \theta) = \frac{1}{\sum\nolimits_{j=1}^K D_j} \sum\limits_{k=1}^K D_k F_k(\bm \theta).
\end{equation}
Following \cite{GuangxuZhu-TWC, YuanmingShi-TWC, ZhibinWang-TWC}, we further assume that the $K$ local datasets have equal size, i.e., $D_k = D$, $\forall k \in {\cal K}$, such that $F(\bm \theta)$ in \eqref{Eq-GLF2} reduces to
\begin{equation}
    F(\bm \theta) = \frac{1}{K} \sum\limits_{k=1}^K F_k(\bm \theta).
\end{equation}

In FL systems, the model parameter vector $\bm \theta$ is trained in a distributed and iterative manner, where the $l$-th training round consists of the following steps.

\textbf{Local gradient computation}: Denote $\bm \theta_k^{[l-1]}$ the local model parameter vector of edge node $k$ at the beginning of the $l$-th training round. To update $\bm \theta_k^{[l-1]}$, edge node $k$ leverages its dataset to compute a local gradient vector, given by
\begin{equation}\label{Eq-LG}
    {\mathbf g}_k^{[l]} ~\triangleq~ \nabla F_k(\bm \theta_k^{[l-1]}) ~=~ \frac{1}{D} \sum\limits_{i=1}^{D} \nabla f(\bm \theta_k^{[l-1]}; \bm \xi_{k, i}, \zeta_{k,i}).
\end{equation}

\textbf{Model aggregation}: As shown in Fig. \ref{MA}, the $K$ edge nodes upload their computed local gradient vectors to the edge server, which takes an average of these local gradient vectors to get the global gradient vector, i.e.,
\begin{equation}\label{Eq-MA}
    \mathbf z^{[l]} = \frac{1}{K} \sum\limits_{k = 1}^K {\mathbf g}_k^{[l]}.
\end{equation}

\textbf{Model dissemination}: As shown in Fig. \ref{MB}, the edge server disseminates $\mathbf z^{[l]}$ to the $K$ edge nodes for local model update.
\begin{equation}\label{Eq-MB}
    \bm \theta_k^{[l]} = \bm \theta_k^{[l-1]} - \eta^{[l]} {\mathbf z}^{[l]},
\end{equation}
where $\eta^{[l]}$ is termed the learning rate.

Such a procedure is repeated for a fixed number of $L$ rounds or until a global consensus is achieved.

\subsection{AirComp-Empowered Model Aggregation}
To reduce communication resource consumption, we adopt AirComp for model uploading. Specifically, at each training round, the $K$ edge nodes upload their respective local gradient vectors to the edge server using the same time-frequency resources. By properly controlling their transmit and receive equalization coefficients and the forwarding matrix of the edge server, a noisy version of the target global gradient vector can be constructed, as detailed below.

First of all, we compute the first-order and second-order statistics of each local gradient vector:
\begin{subequations}
    \begin{eqnarray}
    \bar g_k^{[l]} & = & \frac{1}{d} \sum\limits_{i=1}^d g_{k, i}^{[l]}, \label{First-Order} \\[1ex]
    \delta_k^{[l]} & = & \sqrt{\frac{1}{d} \sum\limits_{i=1}^d \left(g^{[l]}_{k, i} - \bar g_k^{[l]}\right)^2}. \label{Second-Order}
\end{eqnarray}
\end{subequations}
Then, we normalize $g_{k, i}^{[l]}$ using $\bar g_k^{[l]}$ and $\delta_k^{[l]}$, given by
\begin{equation}\label{Eq-Normalization}
    s^{[l]}_{k, i} = \frac{g^{[l]}_{k, i} - \bar g^{[l]}_k}{\delta^{[l]}_k}.
\end{equation}
Through \eqref{Eq-Normalization}, $g^{[l]}_{k, i}$ is normalized as a zero-mean unit-variance symbol $s^{[l]}_{k, i}$, $\forall k \in {\cal K}$.

In uplink model aggregation, we take $s^{[l]}_{k, i}$ as the $i$-th symbol transmitted by edge node $k$, $\forall k \in {\cal K}$. Assuming that the edge server has $N$ antennas, the received signal at the edge server can then be expressed as
\begin{equation}\label{Eq-ServerRxSig}
    \mathbf y_i^{[l]} = \sum\limits_{k = 1}^K {\mathbf h}^{[l]}_{k} b^{[l]}_k s^{[l]}_{k, i} + \mathbf n^{[l]}_{i},
\end{equation}
where $b^{[l]}_{k} \in {\cal C}$ is the transmit equalization coefficient of edge node $k$, $\mathbf h_k \in {\cal C}^{N \times 1}$ is the uplink channel from edge node $k$ to the edge server, $\forall k \in {\cal K}$, and $\mathbf n^{[l]}_{i} \in {\cal C}^{N \times 1}$ is the additive white Gaussian noise at the edge server, which follows ${\cal CN} \left(\mathbf 0, \sigma_s^2 \mathbf I \right)$. Besides, the average power constraint for each edge node is considered such that 
\begin{equation}\label{Eq-PowerConstraint}
    \frac{1}{d} \sum\limits_{i=1}^d \left|b_k^{[l]} s^{[l]}_{k, i} \right|^2 = \left|b^{[l]}_{k}\right|^2 \le P_k,~\forall k \in {\cal K}.
\end{equation}

Upon receiving $\mathbf y_i^{[l]} \in {\cal C}^{N \times 1}$, the edge server processes it using a forwarding matrix $\mathbf M^{[l]} \in {\cal C}^{N \times N}$, i.e.,
\begin{equation}\label{Eq-ServerTxSig}
    \mathbf x_i^{[l]} = \mathbf M^{[l]} \mathbf y_i^{[l]},
\end{equation}
and then disseminates $\mathbf x_i^{[l]} \in {\cal C}^{N \times 1}$ to the $K$ edge nodes with the following power constraint:
\begin{align}\label{Eq-ServerPowerConstraint}
    & \frac{1}{d} \sum\limits_{i = 1}^d {\mathbb E}\left[\left\|\mathbf x_i^{[l]}\right\|^2\right] ~=~ \frac{1}{d} \sum\limits_{i = 1}^d {\mathbb E}\left[\left\|\mathbf M^{[l]} \mathbf y_i^{[l]}\right\|^2\right] \nonumber \\[1ex]
    & ~~~~~~~=~ \frac{1}{d} \sum\limits_{i = 1}^d {\mathbb E} \left[\left\|\mathbf M^{[l]}\left(\mathbf H^{[l]} \mathbf B^{[l]} \mathbf s_i^{[l]} + \mathbf n_i^{[l]}\right)\right\|^2\right] \nonumber \\[2ex]
    & ~~~~~~~=~ {\rm Tr}\left(\mathbf M^{[l]} \mathbf H^{[l]} \mathbf B^{[l]} \mathbf S^{[l]} \mathbf B^{H, [l]} \mathbf H^{H, [l]} \mathbf M^{H, [l]}\right) \nonumber \\[2ex]
    & ~~~~~~~+~~ \sigma_s^2 {\rm Tr}\left(\mathbf M^{[l]} \mathbf M^{H, [l]}\right) \le P_s,
\end{align}
where $\mathbf H^{[l]} = \big[\mathbf h^{[l]}_1, \cdots, \mathbf h^{[l]}_K\big]$, $\mathbf B^{[l]} = {\rm diag}\big\{b^{[l]}_1, \cdots, b^{[l]}_K\big\}$, $\mathbf s_i^{[l]} = \big[s_{1, i}^{[l]}, \cdots, s_{K, i}^{[l]}\big]^T$, and $\mathbf S^{[l]} = \frac{1}{d} \sum\nolimits_{i=1}^d \mathbf s_i^{[l]} \mathbf s_i^{T,[l]}$.

Denoting $\mathbf q_k^{H, [l]} \in {\cal C}^{1 \times N}$ as the downlink channel from edge server to edge node $k$, $\forall k \in \mathcal K$, the received signal of edge node $k$ can then be expressed as
\begin{eqnarray}\label{Eq-NodeRxSig}
    r_{k, i}^{[l]} & = & a_k^{[l]} \left(\mathbf q_k^{H, [l]} \mathbf x_i^{[l]} + n_{k, i}^{[l]}\right) \nonumber \\[2ex]
    & = & a_k^{[l]} \mathbf q_k^{H, [l]} \mathbf M^{[l]} \sum\limits_{j = 1}^K {\mathbf h}^{[l]}_{j} b^{[l]}_j \left(\frac{g^{[l]}_{j, i} - \bar g^{[l]}_j}{\delta^{[l]}_j}\right) \nonumber \\[2ex] 
    & + & a_k^{[l]} \mathbf q_k^{H, [l]} \mathbf M^{[l]} \mathbf n^{[l]}_{i} + a_k^{[l]} n_{k, i}^{[l]},
\end{eqnarray}
where $a_k^{[l]} \in {\cal C}$ is the receive equalization coefficient of edge node $k$, and $n^{[l]}_{k, i} \in {\cal C}$ is the additive white Gaussian noise at this node, which follows ${\cal CN} \left(0, \sigma_k^2\right)$. Note that we consider a block fading channel model in \eqref{Eq-ServerRxSig} and \eqref{Eq-NodeRxSig}, where the channel gain coefficient of each link is assumed to be invariant within one training round, such that both $\big\{\mathbf h_k^{[l]}, b_k^{[l]}\big\}$ and $\big\{\mathbf q_k^{[l]}, a_k^{[l]}\big\}$ are unrelated to $i$. Moreover, following existing literature, e.g., \cite{HangLiu-TWC} and \cite{ZhibinWang-TWC}, we assume that $\big\{\mathbf h_k^{[l]}\big\}$ and $\big\{\mathbf q_k^{[l]}\big\}$ are available at the edge server, which is responsible for the overall system optimization.

Following \cite{YuanmingShi-TWC, YongmingHuang-JSAC, HangLiu-TWC, ZhibinWang-TWC, MIMO-OTA}, we adopt the popular uniform-forcing design to recover a noisy version of the global gradient vector at each edge node, i.e.,
\begin{equation}\label{Eq-UF}
    a_k^{[l]} \mathbf q_k^{H, [l]} \mathbf M^{[l]} \mathbf h_j^{[l]} b_j^{[l]} / \delta_j^{[l]} = 1, ~\forall k, j \in {\cal K}.
\end{equation}

Using \eqref{Eq-UF}, we can reduce \eqref{Eq-NodeRxSig} to  
\begin{equation}\label{Eq-NodeRxSig-M2}
    r_{k, i}^{[l]} = \sum\limits_{j = 1}^K \left(g^{[l]}_{j, i} - \bar g^{[l]}_j\right) + a_k^{[l]} \mathbf q_k^{H, [l]} \mathbf M^{[l]} \mathbf n^{[l]}_{i} + a_k^{[l]} n_{k, i}^{[l]}.
\end{equation}
By first adding $\sum\nolimits_{j = 1}^K \bar g^{[l]}_j$, and then multiplying $\frac{1}{K}$ on both sides of \eqref{Eq-NodeRxSig-M2}, we obtain\footnote{For the sake of simplicity, we follow \cite{HangLiu-TWC} and assume edge node $k$, $\forall k \in {\cal K}$, sends $\bar g_k^{[l]}$ and $\delta_k^{[l]}$ to the edge server in an error-free fashion.}
\begin{eqnarray}\label{Eq-NodeRxSig-M3}
    z_{k, i}^{[l]} ~\triangleq  \frac{1}{K} \left(r^{[l]}_{k,i} + \sum\limits_{k = 1}^K \bar g^{[l]}_k \right) ~~~~~~~~~~~~~~~~~~~~~~~~~~~~~~~~ \nonumber \\[2ex]
    = \frac{1}{K} \sum\limits_{k = 1}^K g^{[l]}_{k, i} + \frac{1}{K}\left(a_k^{[l]} \mathbf q_k^{H, [l]} \mathbf M^{[l]} \mathbf n^{[l]}_{i} + a_k^{[l]} n_{k, i}^{[l]}\right).
\end{eqnarray}
Comparing \eqref{Eq-NodeRxSig-M3} with \eqref{Eq-MA}, it is observed that $z_{k, i}^{[l]}$ is a noisy version of $z_{i}^{[l]}$. In the sequel, we define $e_{k, i}^{[l]}$ as
\begin{eqnarray}
    e_{k, i}^{[l]} ~\triangleq~ z_{k, i}^{[l]} - z_i^{[l]} ~~~~~~~~~~~~~~~~~~~~~~~~~~~~ \nonumber \\[2ex]
    = \frac{1}{K}\left(a_k^{[l]} \mathbf q_k^{H, [l]} \mathbf M^{[l]} \mathbf n^{[l]}_{i} + a_k^{[l]} n_{k, i}^{[l]}\right).
\end{eqnarray}
The received global gradient vector through wireless channels inevitably becomes inaccurate due to fading and noise, leading to a negative effect on the FL learning performance\footnote{In this paper, the two terms ``learning performance'' and ``convergence performance'' are somehow equivalent. Specifically, we define an FL system to have good learning performance when its training loss is close to the optimal loss after some iterations.}, as detailed in Section \ref{Section-Convergence}.

\section{FL Convergence Performance in the Presence of Gradient Error}\label{Section-Convergence}
In this section, we analyze the convergence property of the considered wireless FL system, which motivates the proposed
uplink and downlink transceiver designs in the next section. To proceed, we follow \cite{MingzheChen-TWC, HaoChen-IoTJ, HangLiu-TWC, MPFriedlander} and make the following assumptions.
\begin{assumption}
The global loss function $F(\cdot)$ is uniformly Lipschitz continuous with parameter $\rho > 0$, such that for any $\bm \theta, \bm \theta' \in {\cal R}^d$, we have
\begin{equation}\label{Eq-SmoothAssump}
    F(\bm \theta') \le F(\bm \theta) + \left(\bm \theta' - \bm \theta\right)^{T} \nabla F(\bm \theta) + \frac{\rho}{2} \left\|\bm \theta' - \bm \theta \right\|^2.
\end{equation}
\end{assumption}

\begin{assumption}
The global loss function $F(\cdot)$ is strongly convex with respect to parameter $\mu > 0$, such that for any $\bm \theta, \bm \theta' \in {\cal R}^d$, we have
\begin{equation}\label{Eq-ConvexAssump}
    F(\bm \theta') \ge F(\bm \theta) + \left(\bm \theta' - \bm \theta \right)^{T} \nabla F(\bm \theta) + \frac{\mu}{2} \left\|\bm \theta' - \bm \theta \right\|^2.
\end{equation}
\end{assumption}

Suppose that the global loss function $F\left(\bm \theta\right)$ indeed satisfies the above two assumptions and the learning rate  $\eta^{[l]}$ is set to ${1}/{\rho}$. Following \cite{MPFriedlander}, we can derive that
\begin{equation}\label{Eq-OneIter}
    F\big(\bm \theta_k^{[l+1]}\big) \le F\big(\bm \theta_k^{[l]}\big) - \frac{1}{2\rho}\big\|\nabla F\big(\bm \theta_k^{[l]}\big)\big\|^2 + \frac{1}{2\rho} \big\|\mathbf e_k^{[l]}\big\|^2,
\end{equation}
where $\mathbf e^{[l]}_k = \big[e^{[l]}_{k,1}, \cdots, e^{[l]}_{k, d}\big]^T$, $\forall k \in {\cal K}$. As $\big\|\mathbf e_k^{[l]}\big\|^2$ is unbounded, we turn to its expectation given by
\begin{eqnarray}\label{Eq-CNEB}
    {\mathbb E}\left[\big\|\mathbf e_k^{[l]}\big\|^2\right] = \sum\limits_{i=1}^d {\mathbb E} \left[\big|e^{[l]}_{k, i}\big|^2\right] ~~~~~~~~~~~~~~~~~~~~~~~~~~~~ \nonumber \\[2ex]
    = \frac{d}{K^2} \big|a^{[l]}_k\big|^2 \left(\sigma_s^2 \mathbf q_k^{H, [l]} \mathbf M^{[l]} \mathbf M^{H, [l]} \mathbf q_k^{[l]} + \sigma_k^2 \right).
\end{eqnarray}

Then, based on \eqref{Eq-OneIter} and \eqref{Eq-CNEB}, we can derive the following theorem.
\begin{theorem}\label{Theorem-NSFL-Convergence}
    \emph{Suppose that Assumption 1 and Assumption 2 are valid and the learning rate is fixed to ${1}/{\rho}$. After $L \ge 1$ training rounds, the expected difference between the training loss and the optimal loss using $\bm \theta_k^{[1]}, \cdots, \bm \theta_k^{[L]}$ can be upper bounded by}
    \begin{eqnarray}\label{Eq-NSFL-Convergence}
        {\mathbb E}\left[F\big(\bm \theta_k^{[L+1]}\big) - F\big(\bm \theta^{\star}\big)\right] & \le & {\mathbb E}\left[F\big(\bm \theta_k^{[1]}\big) - F\big(\bm \theta^{\star}\big)\right] ~ \lambda^L \nonumber \\[2ex]
        & + & \sum\limits_{l=1}^{L} \frac{\lambda^{L-l}}{2 \rho} {\mathbb E}\left[\big\|\mathbf e_k^{[l]}\big\|^2\right],
    \end{eqnarray}
    \emph{where $\bm \theta^{\star}$ denotes the optimal model parameter vector and $\lambda = 1 - \mu / \rho$.}
\end{theorem}
\begin{IEEEproof}
    \emph{Refer to Appendix \ref{Theorem-NSFL-Proof}.}
\end{IEEEproof}

Since $\lambda \in (0, 1)$, when $L \rightarrow \infty$, $\lambda^{L} \rightarrow 0$, and we can thus simplify \eqref{Eq-NSFL-Convergence} as
\begin{equation}\label{Eq-NSFL-Convergence2}
    {\mathbb E}\left[F\big(\bm \theta_k^{[L+1]}\big) - F\big(\bm \theta^{\star}\big)\right] ~\le~ \sum\limits_{l = 1}^{L} \frac{\lambda^{L-l}}{2 \rho} {\mathbb E}\left[\big\|\mathbf e_k^{[l]}\big\|^2\right].
\end{equation}
It can be observed from \eqref{Eq-NSFL-Convergence2} that FL recursions over wireless channels still converge, though a gap between $F(\bm \theta^{\star})$ and $\lim\nolimits_{L \to \infty}{\mathbb E}[F(\bm \theta_k^{[L+1]})]$ exists due to communication errors. In the next section, we will jointly optimize each edge node transmit and receive equalization coefficients along with the edge server forwarding matrix to minimize the maximum MSE among the $K$ edge nodes in each training round, aiming to improve the performance of the considered wireless FL system.

\section{Joint Uplink-Downlink Design}\label{Section-NSFL-PF}
To proceed, we focus on the $l$-th training round and take the maximum MSE among the $K$ edge nodes as the objective function to construct the following optimization problem:
\begin{subequations}\label{Eq-OP0}
    \begin{eqnarray}
        (\mathcal P_0) \min\limits_{\mathbf M, \{a_k\}, \{b_k\}} \max\limits_{\forall k \in {\cal K}} \left\{\big|a_k\big|^2 \left(\sigma_s^2 \mathbf q_k^{H} \mathbf M \mathbf M^{H} \mathbf q_k + \sigma_k^2 \right)\right\} \label{OP0-Obj} \\[2ex]
        \text{s.t.}~~~~~~~ a_k \mathbf q_k^{H} \mathbf M \mathbf h_j {b_j}/{\delta_j} = 1, ~\forall k, j \in {\cal K}, ~~~~~ \label{OP0-UF} \\[2ex]
        \eqref{Eq-PowerConstraint},~\eqref{Eq-ServerPowerConstraint}, ~~~~~~~~~~~~~~~~~~~~~~~~~~~~~~~
\end{eqnarray}
\end{subequations}
where the training round index $l$ has been dropped for brevity. Before solving this problem, we first provide the following theorem.

\begin{theorem}\label{theorem-rank}
    \emph{Considering the case of $K \ge N$, to ensure (\ref{OP0-UF}) is satisfied, the rank of $\bm M$ must be equal to one.}
\end{theorem}
\begin{IEEEproof}
    \emph{Refer to Appendix \ref{proof-rank}.}
\end{IEEEproof}

Based on Theorem \ref{theorem-rank}, we can decompose the edge server's forwarding matrix $\mathbf M$ into
\begin{equation}\label{Eq-M}
    \mathbf M = \sqrt{\beta} \mathbf u \mathbf v^{H},
\end{equation}
where $\mathbf u \in {\cal C}^{N \times 1}$ and $\mathbf v \in {\cal C}^{N \times 1}$ are two unit-norm vectors, i.e., $\|\mathbf u\| = \|\mathbf v\| = 1$, and $\beta > 0$ is used to control the transmit power of the edge server.

\begin{corollary}\label{theorem-TxRxDesign}
    \emph{Once $\bm M = \sqrt{\beta} \bm u \bm v^{H}$, to ensure (\ref{OP0-UF}) is satisfied, we have}
    \begin{eqnarray}
        b_k & = & \sqrt{\phi} {\delta_k} \frac{\mathbf h_k^{H} \mathbf v}{\big|\mathbf v^{H} \mathbf h_{k}\big|^2}, ~\forall k \in {\cal K}, \label{Eq-ZF} \\[2ex]
        a_k & = & \frac{1}{\sqrt{\beta \phi}} \frac{\mathbf u^{H} \mathbf q_k}{\big|\mathbf u^{H} \mathbf q_k\big|^2}, ~\forall k \in {\cal K}. \label{Eq-NodeRxFactor}
    \end{eqnarray}
    \emph{where $\phi > 0$ is used to control the transmit power of the $K$ edge nodes.}
\end{corollary}
\begin{IEEEproof}
    \emph{Refer to Appendix \ref{proof-TxRxDesign}.}
\end{IEEEproof}

Given $b_k$ in \eqref{Eq-ZF}, the edge node power constraint in \eqref{Eq-PowerConstraint} can be rewritten as
\begin{equation}\label{Eq-PowerConstraint2}
    \big|b_k\big|^2 = \frac{\delta_k^2 \phi}{|\mathbf v^{H} \mathbf h_k|^2} \le P_k, ~\forall k \in {\cal K}.
\end{equation}
Moreover, by using \eqref{Eq-M}, \eqref{Eq-ZF}, and \eqref{Eq-NodeRxFactor}, we can respectively simplify $\big|a_k\big|^2 \left(\sigma_s^2 \mathbf q_k^{H} \mathbf M \mathbf M^{H} \mathbf q_k + \sigma_k^2 \right)$ and $ \allowdisplaybreaks \sigma_s^2 {\rm Tr}(\mathbf M \mathbf M^H) + {\rm Tr}(\mathbf M \mathbf H \mathbf B \mathbf S \mathbf B^H \mathbf H^H \mathbf M^H )$ as
\begin{eqnarray}\label{Eq-MSE2}
    \big|a_k\big|^2 \left(\sigma_s^2 \mathbf q_k^{H} \mathbf M \mathbf M^{H} \mathbf q_k + \sigma_k^2 \right) = \frac{\sigma_s^2}{\phi} + \frac{\sigma_k^2}{\beta \phi |\mathbf q_k^{H} \mathbf u|^2},
\end{eqnarray}
\begin{eqnarray}\label{Eq-ServerPowerConstraint2}
    {\rm Tr}(\mathbf M \mathbf H \mathbf B \mathbf S \mathbf B^H \mathbf H^H \mathbf M^H ) + \sigma_s^2 {\rm Tr}(\mathbf M \mathbf M^H) ~~~~~~~~~ \nonumber \\[2ex]
    =~ \beta \phi c + \beta \sigma_s^2,
\end{eqnarray}
where $c = \bm \delta^{T} \mathbf S \bm \delta$, and $\bm \delta = \big[\delta_1, \cdots, \delta_K\big]^T$. Based on \eqref{Eq-PowerConstraint2}, \eqref{Eq-MSE2}, and \eqref{Eq-ServerPowerConstraint2}, we equivalently transform $\mathcal P_0$ into
\begin{subequations}\label{Eq-OP1}
    \begin{eqnarray}
        \hspace{-1.25cm} ({\mathcal P}_1) & \min\limits_{\beta, \phi, \mathbf u, \mathbf v} & \max\limits_{\forall k \in {\cal K}} \left\{\frac{\sigma_s^2}{\phi} + \frac{\sigma_k^2}{\beta \phi |\mathbf q_k^{H} \mathbf u|^2}\right\} \label{OP1-Obj} \\[2ex]
        \hspace{-1.25cm} & \text{s.t.} & \frac{\delta^2_k \phi}{|\mathbf v^{H} \mathbf h_{k}|^2} \le P_k, ~\forall k \in {\cal K}, \label{OP1-NodePowerCons} \\[2ex]
        \hspace{-1.25cm} & & \beta \phi c + \beta \sigma_s^2 \le P_s, \label{OP1-ServerPowerCons} \\[2ex]
        \hspace{-1.25cm} & & \|\mathbf u\|^2 = \|\mathbf v\|^2 = 1. \label{OP1-NormCons}
\end{eqnarray}
\end{subequations}
Since increasing $\beta$ leads to a decrease of the objective function \eqref{OP1-Obj}, we can thus replace \eqref{OP1-ServerPowerCons} with $\beta \phi c + \beta \sigma_s^2 = P_s$. That is, the optimal $\beta$ to $\mathcal P_1$ is given by
\begin{equation}\label{Eq-Optibeta}
    \beta^{\star} = \frac{P_s}{\phi c + \sigma_s^2}.
\end{equation}
Next, by substituting \eqref{Eq-Optibeta} into \eqref{Eq-OP1}, we eliminate $\beta$ and reformulate $\mathcal P_1$ into a problem on $\phi$, $\mathbf u$, and $\mathbf v$ only, given by   
\begin{subequations}\label{Eq-OP2}
    \begin{eqnarray}
        \hspace{-0.5cm} (\mathcal P_2) & \min\limits_{\phi, \mathbf u, \mathbf v} & \frac{\sigma_s^2}{\phi} + \max\limits_{\forall k \in {\cal K}} \left\{\frac{\sigma_k^2}{|\mathbf q_k^{H} \mathbf u|^2 P_s}\right\} \left(c + \frac{\sigma_s^2}{\phi}\right)  \\[2ex]
        \hspace{-0.5cm} & \text{s.t.} & \frac{\delta^2_k \phi}{|\mathbf v^{H} \mathbf h_{k}|^2} \le P_k, ~\forall k \in {\cal K}, \\[3ex]
        \hspace{-0.5cm} & & \|\mathbf u\|^2 = \|\mathbf v\|^2 = 1.
\end{eqnarray}
\end{subequations}
It can be observed from \eqref{Eq-OP2} that $\mathbf u$ and $\{\phi, \mathbf v\}$ are decoupled. Therefore, we decompose $\mathcal P_2$ into two subproblems and optimize $\mathbf u$ and $\{\phi, \mathbf v\}$ separately. The subproblem associated with $\mathbf u$ is formulated as follows
\begin{subequations}\label{Eq-OP31}
    \begin{eqnarray}
        \hspace{-2.0cm} (\mathcal P_{3-1}) & \min\limits_{\mathbf u} & \max\limits_{\forall k \in {\cal K}} \left\{\frac{\sigma_k^2}{|\mathbf q_k^{H} \mathbf u|^2}\right\}  \\[2ex]
        \hspace{-2.0cm} & \text{s.t.} & \|\mathbf u\|^2 = 1,
\end{eqnarray}
\end{subequations}
which is equivalent to the following optimization problem
\begin{subequations}\label{Eq-OP32}
    \begin{eqnarray}
        & \min\limits_{\mathbf u, \Lambda} &  \Lambda \\
        & \text{s.t.} & \frac{\sigma_k^2}{|\mathbf q_k^{H} \mathbf u|^2} \le \Lambda, ~\forall k \in {\cal K}, \label{OP32-Cons1} \\[2ex]
        & & \|\mathbf u\|^2 = 1, \label{OP32-Cons2}
\end{eqnarray}
\end{subequations}
where $\Lambda > 0$ is an auxiliary variable. To cope with the non-convexity of \eqref{OP32-Cons1} and \eqref{OP32-Cons2}, we leverage the matrix lifting technique by defining $\mathbf U \triangleq \mathbf u \mathbf u^H$, and transform \eqref{Eq-OP32} into
\begin{subequations}\label{Eq-OP4}
    \begin{eqnarray}
        & \min\limits_{\mathbf U, \Lambda} &  \Lambda \label{OP4-Obj} \\
        & \text{s.t.} & \frac{\sigma_k^2}{\Lambda} \le {\rm Tr}(\mathbf q_k \mathbf q_k^H \mathbf U), ~\forall k \in {\cal K}, \label{OP4-Cons1} \\[2ex]
        & & {\rm Tr}(\mathbf U) = 1, \label{OP4-Cons2} \\[2ex]
        & & {\rm Rank}(\mathbf U) = 1. \label{OP4-Cons3}
    \end{eqnarray}
\end{subequations}
Regarding the non-convex constraint \eqref{OP4-Cons3}, note that it is equivalent to ${\rm Tr}(\mathbf U) - \|\mathbf U\| = 0$. We include ${\rm Tr}(\mathbf U) - \|\mathbf U\|$ as a penalty term to \eqref{OP4-Obj} and transform \eqref{Eq-OP4} into 
\begin{subequations}\label{Eq-OP5}
    \begin{eqnarray}
        \hspace{-1cm} & \min\limits_{\mathbf U, \Lambda} &  \Lambda + \Omega ({\rm Tr}(\mathbf U) - \|\mathbf U\|) \label{OP5-Obj} \\[1ex]
        \hspace{-1cm} & \text{s.t.} & \eqref{OP4-Cons1}, ~\eqref{OP4-Cons2},
    \end{eqnarray}
\end{subequations}
where $\Omega > 0$ is a tuning parameter. While \eqref{OP5-Obj} is still non-convex, its structure of minimizing the difference between two convex functions can be leveraged to develop efficient DC algorithms. In what follows, we use the SCA technique to solve \eqref{Eq-OP5}. Specifically, at iteration $n + 1$, by linearizing the concave parts in \eqref{OP5-Obj}, i.e.,
\begin{eqnarray*}
    \|\mathbf U\| & \ge & \|\mathbf U_n\| + \langle \partial_{\mathbf U_n} \|\mathbf U\|, \mathbf U - {\mathbf U_n} \rangle = {\rm Tr}(\bm \omega_n \bm \omega_n^{H} \mathbf U),
\end{eqnarray*}
we can construct a convex optimization problem given by
\begin{subequations}\label{Eq-OP6}
    \begin{eqnarray}
        & \min\limits_{\mathbf U, \Lambda} & \Lambda + \Omega ({\rm Tr}(\mathbf I - \bm \omega_n \bm \omega_n^{H}) \mathbf U) \\[1ex]
        & \text{s.t.} & \eqref{OP4-Cons1}, ~\eqref{OP4-Cons2}.
\end{eqnarray}
\end{subequations}
where $\partial_{\mathbf U_n} \|\mathbf U\|$ denotes the sub-gradient of $\|\mathbf U\|$ at $\mathbf U_n$, and $\bm \omega_n$ is the eigenvector associated with the largest eigenvalue of $\mathbf U_n$. According to \cite{YuanmingShi-TWC}, we have $\partial_{\mathbf U_n} \|\mathbf U\| = \bm \omega_n \bm \omega_n^{H}$. Solving \eqref{Eq-OP6} successively until convergence, we can obtain a rank-one $\mathbf U$, denoted by $\mathbf U^{\star}$. We then extract $\mathbf u^{\star}$ by doing Cholesky decomposition for $\mathbf U^{\star}$.

Until now, we have introduced how to optimize $\mathbf u$. Below we introduce the optimization of $\mathbf v$ and $\phi$. Their associated subproblem is given by
\begin{subequations}\label{Eq-OP7}
    \begin{eqnarray}
        (\mathcal P_{3-2}) & \min\limits_{\phi, \mathbf v} & \frac{\sigma_s^2}{\phi}  \\[1ex]
        & \text{s.t.} & \frac{\delta^2_k}{|\mathbf v^{H} \mathbf h_{k}|^2} \le \frac{P_k}{\phi}, ~\forall k \in {\cal K}, \\[2ex]
        & & \|\mathbf v\|^2 = 1.
\end{eqnarray}
\end{subequations}
It can be observed that \eqref{Eq-OP7} shares almost the same form as \eqref{Eq-OP32} except for replacing $\mathbf u$ and $\Lambda$ with $\mathbf v$ and ${1}/{\phi}$, and hence the techniques for solving \eqref{Eq-OP32}, i.e., matrix lifting and DC transformation, can be used to solve \eqref{Eq-OP7} as well. Here we omit the details for brevity.

\section{Numerical Results}\label{Section-NR}
We consider a three-dimensional coordinate system, where the location of the edge server is set to $\left(-50, 0, 10\right)$ meters, and the $K = 20$ edge nodes are uniformly distributed in the region of $\left([0,20], [-10, 10], 0\right)$ meters. The uplink and downlink channels between edge node $k$ and the edge server, i.e., $\mathbf h_{k}$, and $\mathbf q^H_k$, $\forall k \in {\cal K}$, suffer from both path loss and small-scale fading \cite{ZhibinWang-TWC}. The path loss model is expressed as $\textsf{PL}(\Gamma) = C_0 \left(\Gamma / \Gamma_0\right)^{-\kappa}$, where $C_0 = 30$ dB accounts for the path loss at the reference distance of $\Gamma_0 = 1$ meter, $\Gamma$ denotes the link distance, and $\kappa = 2.2$ is the path loss component. The small-scale fading is modeled as
\begin{equation*}
    \sqrt{\frac{\chi}{1 + \chi}} + \sqrt{\frac{1}{1 + \chi}}{\cal CN}(0, 1),
\end{equation*}
where $\chi = 1$ is termed the Rician factor. Moreover, we set $\sigma_s^2 = -50$ dBW, $\sigma_k^2 = -50$ dBW, and $P_k = 0$ dBW, $\forall k \in {\cal K}$.

Regarding the learning purpose, we use the MNIST dataset \cite{MNIST} to simulate the handwritten digit recognition task. Specifically, by using cross-entropy as the loss function, we train a fully connected neural network consisting of 784 inputs and 10 outputs, i.e., the number of model parameters $d = 7840$. The training set of 60,000 samples is equally divided into 20 shards of size $D = 3000$ in a non-IID manner, and each shard is assigned to one edge node as its local dataset. The test dataset has 10,000 different samples, and we adopt test accuracy, defined as $\frac{\text{\# of correctly recognized handwritten-digits}}{10000} \in [0,1]$, to evaluate the FL learning performance. The total number of training rounds $L = 50$, and the learning rate $\eta^{[l]}$ is set to $ 0.01$, $\forall l = 1, \cdots, L$.

\begin{figure}
\vskip -8pt
\centering
\subfigure[Average MSE of the $K$ edge nodes.]
{\includegraphics[width = 7.8cm]{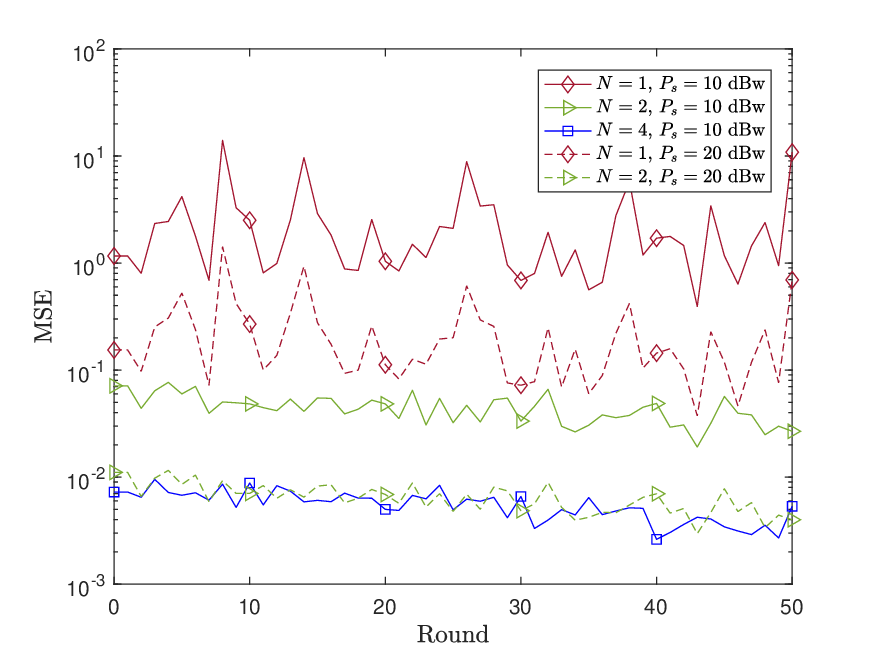} \label{MSE1-Round}}
$~$\subfigure[Average recognition accuracy of the $K$ edge nodes.]
{\includegraphics[width = 7.8cm]{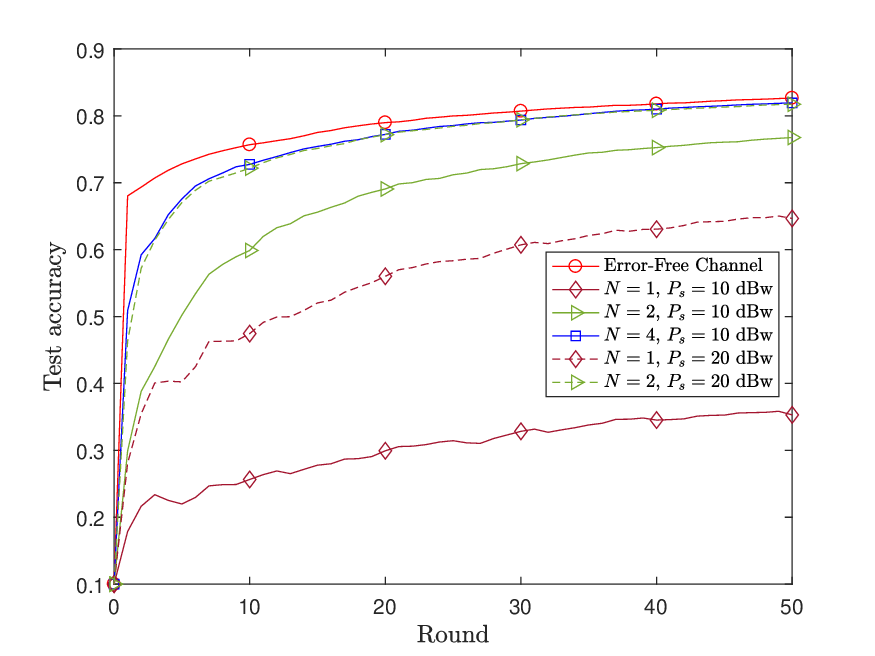} \label{ACC1-Round}}
\caption{Performance with respect to the number of FL training rounds.}\vspace{-1em}
\end{figure}

In Fig. \ref{MSE1-Round}, we plot the average MSE of the $K$ edge nodes, defined as $\frac{1}{Kd}\sum\nolimits_{k = 1}^K {\mathbb E}\big[\|\mathbf e_k^{[l]}\|^2\big]$, $\forall l = 1, \cdots, L$. From this figure, we immediately observe that the average MSE of the $K$ edge nodes drops significantly when we increase the number of antennas at the edge server, demonstrating the great potential of introducing multiple antennas at the edge server. Moreover, by increasing the forwarding power of the edge server from $10$ dBw to $20$ dBw, the average MSE also drops notably. We can also observe from Fig. \ref{MSE1-Round} that the curve of the average MSE becomes smoother when the number of antennas at the edge server is increased, exhibiting the ``channel hardening'' effect. Furthermore, as shown in Fig. \ref{ACC1-Round}, the recognition accuracy improves significantly by increasing the number of antennas at the edge server from $N = 1$ to $N = 4$ or by increasing $P_s$ from $10$ dBw to $20$ dBw, which can be attributed to the decreasing MSE and thus corroborates Theorem \ref{Theorem-NSFL-Convergence}.

\section{Conclusions}\label{Section-CN}
In this paper, we focused on an AirComp-empowered FL system and studied its communication designs by considering uplink model aggregation and downlink model dissemination jointly, which was different from most existing works. We analyzed the convergence performance of the considered FL system, demonstrating that it was related to the MSE between the target global gradient vector and the received one at each edge node. Accordingly, we further optimized each edge node transmit and receive equalization coefficients along with the edge server forwarding matrix to minimize the maximum MSE across all edge nodes. The performance of the considered FL system was also evaluated through the handwritten digit recognition task. Experiment results verified that deploying multiple antennas at the edge server can significantly reduce the MSE at each edge node, leading to a remarkable improvement in the recognition accuracy compared to the single antenna case.

\begin{appendices}
\section{}\label{Theorem-NSFL-Proof}
It can be derived from \eqref{Eq-ConvexAssump} that
\begin{eqnarray}\label{Eq-Grad-LB}
    \big\|\nabla F\big(\bm \theta_k^{[l]}\big)\big\|^2 & \ge & 2 \mu \big[F\big(\bm \theta_k^{[l]}\big) - F\big(\bm \theta^{\star}\big)\big].
\end{eqnarray}
Then, by substituting \eqref{Eq-Grad-LB} into \eqref{Eq-OneIter}, we obtain that
\begin{eqnarray}\label{Eq-OneIter2}
    F\big(\bm \theta_k^{[l+1]}\big) ~\le~ F\big(\bm \theta_k^{[l]}\big) - \frac{1}{2\rho} \big\|\nabla F(\bm \theta_k^{[l]})\big\|^2 + \frac{1}{2\rho} \big\|\mathbf e_k^{[l]}\big\|^2 \nonumber \\[2ex]
    \le~ F\big(\bm \theta_k^{[l]}\big) - \frac{\mu}{\rho} \left[F\big(\bm \theta_k^{[l]}\big) - F\big(\bm \theta^{\star}\big)\right] + \frac{1}{2\rho} \big\|\mathbf e_k^{[l]}\big\|^2.
\end{eqnarray}
By first subtracting $F(\bm \theta^{\star})$ and then taking expectation on both sides of \eqref{Eq-OneIter2}, we have
\begin{eqnarray}\label{Eq-OneIter3}
    {\mathbb E}\left[F\big(\bm \theta_k^{[l+1]}\big) - F\big(\bm \theta^{\star}\big)\right] ~~~~~~~~~~~~~~~~~~~~~~~~~~~~~~~~~~~ \nonumber \\[2ex]
    \le~ \lambda~{\mathbb E}\left[F\big(\bm \theta_k^{[l]}\big) - F\big(\bm \theta^{\star}\big)\right] + \frac{1}{2\rho} {\mathbb E}\left[\big\|\mathbf e_k^{[l]}\big\|^2\right].
\end{eqnarray}
Applying \eqref{Eq-OneIter3} recursively for $l = L, \cdots, 1$, we can obtain \eqref{Eq-NSFL-Convergence} and complete the proof.

\section{}\label{proof-rank}
We rewrite \eqref{OP0-UF} in a matrix form, given by
\begin{equation}\label{Eq-A1}
    \mathbf A \mathbf Q^H \mathbf M \mathbf H \mathbf B \bm \Delta^{-1} = \bm \Pi,
\end{equation}
where $\bm \Pi$ is the all-one matrix, $\mathbf A = {\rm diag}\{a_1, \cdots, a_K\}$, $\bm \Delta^{-1} = {\rm diag}\{\delta^{-1}_1, \cdots, \delta^{-1}_K\}$, and $\mathbf Q = [\mathbf q_1, \cdots, \mathbf q_K]$. When $K \ge N$, ${\rm Rank}(\mathbf Q) = {\rm Rank}(\mathbf H) = N$, and we can thus derive that
\begin{eqnarray}
    1 ~=~ {\rm Rank}(\bm \Pi) & = & {\rm Rank}(\mathbf A \mathbf Q^H \mathbf M \mathbf H \mathbf B \bm \Delta^{-1}) \nonumber \\[2ex]
    & \overset{(a)}{=} & {\rm Rank}(\mathbf A \mathbf Q^H \mathbf M) \nonumber \\[2ex]
    & \overset{(b)}{=} & {\rm Rank}(\mathbf M),
\end{eqnarray}
where $(a)$ is due to the property of ${\rm Rank}(\mathbf X \mathbf Y) = {\rm Rank}(\mathbf X)$ for full row rank matrix $\mathbf Y$, and $(b)$ is due to the property of ${\rm Rank}(\mathbf X \mathbf Y) = {\rm Rank}(\mathbf Y)$ for full column rank matrix $\mathbf X$.

\section{}\label{proof-TxRxDesign}
By substituting $\mathbf M = \sqrt{\beta} \mathbf u \mathbf v^H$ into \eqref{OP0-UF}, we obtain that
\begin{equation}\label{Eq-D0}
    \sqrt{\beta} a_k \mathbf q_k^{H} \mathbf u \mathbf v^H \mathbf h_j {b_j}/{\delta_j} = 1,~\forall k, j \in {\cal K}.
\end{equation}
According to \eqref{Eq-D0}, we can then prove $\mathbf v^H \mathbf h_1 b_1/\delta_1 = \cdots = \mathbf v^H \mathbf h_K b_K/\delta_K$, and $\sqrt{\beta} a_1 \mathbf q_1^H \mathbf u = \cdots = \sqrt{\beta} a_K \mathbf q_K^H \mathbf u$, as detailed below.

By introducing $\tilde a_k \triangleq \sqrt{\beta} a_k \mathbf q_k^{H} \mathbf u$, $\forall k \in {\cal K}$, and $\tilde b_j \triangleq \mathbf v^{H} \mathbf h_j {b_j}/{\delta_j}$, $\forall j \in {\cal K}$, we can equivalently rewrite \eqref{Eq-D0} as $\tilde a_k \tilde b_j = 1$, $\forall k, j \in {\cal K}$. Thus, for any $\tilde a_k$, we have
\begin{equation}\label{Eq-D1}
    \tilde a_k = \frac{\tilde b^*_j}{\big|\tilde b_j\big|^2}, ~\big|\tilde a_k\big|^2 = \frac{1}{\big|\tilde b_j\big|^2}, \forall j \in {\cal K}.
\end{equation}
Thanks to \eqref{Eq-D1}, we prove $\tilde b_1, \cdots, \tilde b_K$ have the same phase and amplitude. In other words, we prove that $\tilde b_1 = \cdots = \tilde b_K$. Similarly, we can prove $\tilde a_1 = \cdots = \tilde a_K$ as well.

Consequently, by setting $\mathbf v^H \mathbf h_j b_j/\delta_j = \sqrt{\phi}$, $\forall j \in {\cal K}$, we have $\sqrt{\beta} a_k \mathbf q_k^H \mathbf u = \frac{1}{\sqrt{\phi}}$, $\forall k \in {\cal K}$, which proves Corollary \ref{theorem-TxRxDesign}.
\end{appendices}

\end{document}